\documentclass[aps,twocolumn,prl,superscriptaddress,preprintnumbers,showpacs,tightenlines]{revtex4}
\usepackage{epsfig,graphicx,times}

\def\<{\langle}
\def\>{\rangle}

\begin{document}

\preprint{First Draft}

\title{Perfect Function Transfer in two- and three- dimensions without
  initialization}

\author{Lian-Ao Wu}
\affiliation{Department of Theoretical Physics and History of Science, The Basque Country
University (EHU/UPV), PO Box 644, 48080 Bilbao, Spain}
\affiliation{IKERBASQUE, Basque Foundation for Science, 48011 Bilbao, Spain }
\author{Mark Byrd}
\affiliation{Physics Department, Southern Illinois University, Carbondale, Illinois
62901-4401}
\affiliation{Computer Science Department, Southern Illinois University, Carbondale, Illinois
62901}
\author{Z.-D. Wang}
\affiliation{Department of Physics and Center of Theoretical and Computational Physics,
University of Hong Kong, Pokfulam Road, Hong Kong, China}
\author{Bin Shao}
\affiliation{Department of Physics, Beijing Institute of Technology, Beijing, 100081}
\date{\today }

\begin{abstract}
We find analytic models that can perfectly transfer, without state initialization
or remote collaboration, arbitrary functions in two- and three-dimensional
interacting bosonic and fermionic networks. We elaborate on a possible
implementation of state transfer through bosonic or fermionic atoms trapped
in optical lattices.  A significant finding is that the state of a spin
qubit can be perfectly transferred through a fermionic
system. Families of Hamiltonians, both linear and nonlinear, are described
which are related to the linear Boson model and that enable the
perfect transfer of arbitrary functions.  This includes entangled
states such as decoherence-free subsystems enabling noise protection
of the transferred state.  
\end{abstract}

\maketitle

\pagenumbering{arabic}


\emph{Introduction.---} The study of the transfer of a quantum state
using \emph{naturally-available} interactions in a spin chain dates back to
Bose's work \cite{Bose:03}. The motivation is to enable transfer over
short distances within a quantum information processing system.  
However in most realistic cases it failed to 
perfectly perform such a transfer.  In 2004, Chirstandl 
\emph{et al.}~\cite{Christandl/etal:04} 
found that perfect state transfer (PST) is possible in spin-1/2
networks if parameters in the system's Hamiltonian is carefully
engineered.  Since that time, several general 
results have been obtained for state transfer.  For example, the
fidelity can made arbitrarily large if there is no limit to the 
number of sequential gates that can be applied \cite{Burgarth/etal:06}. 
The communication can also be 
improved if the sender encodes the message 
state over a set of multiple spins 
\cite{Osborne/Linden:04,Haselgrove:05,Allcock:09,Wang/etal:09b},
and ideal state transmissions have also been proposed using quantum dots 
\cite{Petrosyan:04a,Petrosyan:04b}.  (See also the
review~\cite{Bose:07} and references therein.)  
The possibility of perfect transfer triggered some general results
(\cite{Bose:07}, and references therein), \cite{Wu091} and interest in finding
pre-engineered models that analytically demonstrate perfect state
transfer, for example, refs.~\cite{DiFranco} and ~\cite{Sun}.

The strategy for PST in 
Ref.~\cite{Christandl/etal:04,Christandl/etal:05} is to map indices of
the number of sites of a spin chain onto the magnetic quantum numbers
of an angular momentum operator. 
The nearest-neighbor interaction thus becomes the $x$-component
of a quasi-angular momentum. It was recently pointed out that the 
nearest-neighbor interaction can correspond to either the bosonic or spinless
fermionic representations of quasi-angular momenta operators in one dimension \cite%
{Wu09}. As with any analytical solvable model in physics, the simplicity and
beauty of the result clarifies the physical picture and this one
provides \emph{the example} of 
" perfect state transfer."  On the other hand, it is obvious that the
one-dimensional state transfer surely cannot be the end of the story 
of state transmissions. This prompts the following question, can 
the strategy in Ref.~\cite{Christandl/etal:04} be applied to the two- and
three-dimensional cases?  Here, we propose an analytical solution to
the problem of perfectly transferring an unknown function (termed
perfect function transfer or PFT) with one 
or more variables from a processor at one site to another processor at
another site in systems which are \emph{two-} or \emph{three-}dimensional.
The interaction required for perfect state transfer in our model can be
implemented using the Bose-Hubbard model (see, e.g., \cite{Bruder} and
references therein) or fermions in an optical lattice~\cite{Duan}. In
these systems there is a {\it family of Hamiltonians that can perform
  PFT}, obtained via the so-called \emph{dressing} transformations.

\emph{Linear Bose model.---}  Consider the dynamics of cold bosonic
atoms loaded in a two-dimensional square optical lattice with $N$
sites along the horizontal direction and $M$ site along vertical
direction governed by the general linear nearest-neighbor Bose
model Hamiltonian~\cite{Bruder,Jaksch98,Greiner02,Reslen}:  
\begin{equation}
H=-\sum_{j=1}^{M-1}\sum_{k=1}^{N-1}[J_{jk}^{(1)}b_{j,k}^{\dagger
}b_{j+1,k}+J_{jk}^{(2)}b_{j,k}^{\dagger }b_{j,k+1}+\mathrm{h.c.}],
\label{BHH}
\end{equation}%
where $b_{j,k}^{\dagger }$ ($b_{j,k}$) is the bosonic creation
(annihilation) operator at the two-dimensional site $(j,k)$.  
Equation~(\ref{BHH}) describes hopping bosons in the absence of
on-site repulsion. The hopping (or tunneling) matrix element along the
horizontal and vertical directions between nearest-neighbor sites may
be given by 
\begin{eqnarray}
J_{jk}^{(1)}\!\! &=&\!\!\int d^{3}\vec{r}\, w^{\ast }(\vec{r}-\vec{r}_{j,k})
[T+V_{\text{lat}}(\vec{r})]w(\vec{r}-\vec{r}_{j+1,k}), \label{BHH2}\\
J_{jk}^{(2)}\!\! &=&\!\!\int d^{3}\vec{r}\, w^{\ast }(\vec{r}-\vec{r}_{j,k}) 
[T+V_{\text{lat}}(\vec{r})]w(\vec{r}-\vec{r}_{j,k+1}),
\end{eqnarray}
where $w(\vec{r}-\vec{r}_{j,k})$ is a single atom Wannier function at lattice
site $(j,k)$, $V_{\text{lat}}(\vec{r})$ denotes the optical lattice
potential, and $T$ is the kinetic energy of a single atom.

We will consider the limit when the on-site repulsion is neglectably small or the
cases with total boson number being one. The Hamiltonian in Eq.~(\ref{BHH})
is also equivalent to that of the on-chip coupled cavities (e.g., in Refs.~%
\cite{Hartmann,Bliokh}). The tunable coupling $J_{jk}$ between each pair of
resonators can be realized by SQUID couplers~(e.g.,
Ref.~\cite{Sandberg}).  


\emph{Angular momentum and engineered linear Bose model.---} By 
generalizing Ref.~\cite{Christandl/etal:04} 
the indices of the number of sites of a
two-dimensional lattice can be mapped into the magnetic quantum
numbers of the two total 
angular momenta $l_{1},l_{2}$ such that 
$l_{1}=\frac{M-1}{2},l_{2}=\frac{N-1}{2}$ and 
$m_{1}=-\frac{M+1}{2}+j,m_{2}=-\frac{N+1}{2}+k.$ 
For instance, the magnetic numbers for the first site are
$(-\frac{M-1}{2},-\frac{N-1}{2}).$  

With this mapping, the bosonic operator (or spinless fermionic
operator) $b_{j,k}^{\dagger }$ at site $j=m_{1}+\frac{M+1}{2}$ and
$k=m_{2}+\frac{N+1}{2}$ corresponds to the $SU_{1}(2)\times SU_{2}(2)$
irreducible spherical tensor bosonic operator 
$A_{l_{1}m_{1},l_{2}m_{2}}^{\dagger}$, or in a short-hand notation 
$A_{m_{1},m_{2}}^{\dagger }.$ The three components of the horizontal 
(vertical) quasi-angular momentum vector $\vec{L}^{(1)}$ 
($\vec{L}^{(2)}$) can be expressed in terms of the atomic creation and
annihilation operators as 
\begin{eqnarray}
L_{x}^{(1)} &=&\sum_{jk}C_{j}^{(1)}(b_{j,k}^{\dagger
}b_{j+1,k}+b_{j+1,k}^{\dagger }b_{j,k}),\text{ \ }  \nonumber \\
L_{x}^{(2)} &=&\sum_{ik}C_{k}^{(2)}(b_{j,k}^{\dagger
}b_{j,k+1}+b_{j,k+1}^{\dagger }b_{j,k}), \\
L_{y}^{(1)} &=&i\sum_{jk}C_{j}^{(1)}(b_{j,k}^{\dagger
}b_{j+1,k}-b_{j+1,k}^{\dagger }b_{j,k}),  \nonumber \\
L_{y}^{(2)} &=&i\sum_{ik}C_{k}^{(2)}(b_{j,k}^{\dagger
}b_{j,k+1}-b_{j,k+1}^{\dagger }b_{j,k}), \\
L_{z}^{(1)} &=&\sum n_{j,k}\left[ j-{\frac{1}{2}}(M+1)\right] ,  \label{N11}
\\
L_{z}^{(2)} &=&\sum n_{j,k}\left[ k-{\frac{1}{2}}(N+1)\right]
\end{eqnarray}%
where $C_{j}^{(1)}=\frac{1}{2}\sqrt{j(M-j)}$ and 
$C_{k}^{(2)}=\frac{1}{2}\sqrt{k(N-k)}$.

For the tensor product structure of the two $SU(2)$ groups to hold, it
is essential that $\vec{L}^{(1)}$ and $\vec{L}^{(2)}$, 
generators of the groups $SU_{1}(2)$ and $SU_{2}(2)$ commute.  
It is not obvious, but $[\vec{L}^{(1)}$, $\vec{L}^{(2)}]=0$ \emph{does}
hold for bosons {\it and} fermions.  (Fermions will be discussed later.)  
This can be seen by directly calculating the commutators of different
components.  However, unlike in one dimension, the Jordon-Wigner
transformation cannot be used to directly 
replace a fermionic operator by corresponding Pauli matrices $\sigma
_{j,k}^{+}$.  If the $b_{j,k}^{\dagger }\Leftrightarrow
\sigma _{j,k}^{+}$ in (\ref{BHH}), then the corresponding
$\vec{L}^{(1)}$ and $\vec{L}^{(2)}$ will not commute.

Supposing that the $J_{j,k}$ are pre-engineered in Eq.~(\ref{BHH}) as
in Ref.~\cite{Christandl/etal:04} such that 
$J_{j,k}^{(1)}=J^{(1)}C_{j}^{(1)}$ and $J_{j,k}^{(2)}=J^{(2)}C_{k}^{(2)},$
then the time evolution becomes 
\begin{equation}
U(t)=\exp [i(J^{(1)}L_{x}^{(1)}+J^{(2)}L_{x}^{(2)})t].  \label{U(t)}
\end{equation}%
Note that the parameters $J_{j,k}^{(1)}=J^{(1)}C_{j}^{(1)}$ are
constructed such that they are independent of $k$.  

The irreducible tensor operator $A_{m_{1},m_{2}}^{\dagger }$ in the
Heisenberg representation evolves as 
\begin{eqnarray}
U^{\dagger }(t)A_{m_{1},m_{2}}^{\dagger }U(t)&=& 
\sum e^{i\frac{\pi }{2}(m_{1}^{\prime }-m_{1})}d_{m_{1}^{\prime
}m_{1}}^{l_{1}}(J^{(1)}t) \\
&&\times e^{i\frac{\pi }{2}(m_{2}^{\prime
}-m_{2})}d_{m_{2}^{\prime }m_{2}}^{l_{2}}(J^{(2)}t)A_{m_{1}^{\prime
},m_{2}^{\prime }}^{\dagger }
\label{Ut}
\end{eqnarray}%
where $d_{m^{\prime }m}^{l}$ is the small Wigner D-function. When 
$J^{(1)}=J^{(2)}=J$ and $t_{0}=\pi /J,$ this expression reduces to a simple
form: 
\begin{equation}
U^{\dagger}(t_{0})\;b_{j,k}^{\dagger}\;U(t_{0})
      = r_{1}r_{2}b_{M-j+1,N-k+1}^{\dagger}\,,  
\label{transfer}
\end{equation}%
where the factors $r_{1}=\exp \left( -i\pi \frac{M-1}{2}\right)$ and $%
r_{2}=\exp \left( -i\pi \frac{N-1}{2}\right) $ are analogous to the 
\emph{signature} in nuclear structure theory \cite{Ring:book}.  These
factors determine the interference of quantum states traversing the
chain and depend on the number of lattice sites \cite{Wu09}. 
Choosing the number of sites appropriately can produce interference 
such that $r_{1}r_{2}=1.$  Along with the two directions after $t_0$,
we have  
\begin{eqnarray*}
e^{iL_{x}^{(1)}\pi }b_{j,k}^{\dagger }e^{-iL_{x}^{(1)}\pi }
&=&r_{1}b_{M-j+1,k}^{\dagger } \\
e^{iL_{x}^{(2)}\pi }b_{j,k}^{\dagger }e^{-iL_{x}^{(2)}\pi }
&=&r_{2}b_{j,N-k+1}^{\dagger }
\end{eqnarray*}%
which reduces the problem to one-dimensional chains.  

\emph{Perfect state transfer.---} Let us first consider the case of a
bosonic system which can be applied to various linear optical or
atomic systems.  Suppose that a known
or unknown function $f$ is encoded in the bottom-left site $(1,1)$, such
that $f(x)$ is mapped to the state 
$f(b_{1,1}^{\dagger})|\mathbf{0}\rangle$, where 
$|\mathbf{0}\rangle =|0\rangle^{\otimes  N}$. 
Thus, a general function can be perfectly
transferred to the top-right $(M,N)$ by 
\begin{equation}
U(t_{0})f(b_{1,1}^{\dagger })|\mathbf{0}\rangle
=f(b_{M,N}^{\dagger })|\mathbf{0}\rangle.  \label{N5}
\end{equation}
We emphasize that the central sites, other than $(1,1)$ and $(M,N)$, of the
two-dimensional network are not necessarily in the ground state and can be in
an arbitrary state $g(b_{j,k}^{\dagger })|\mathbf{0}\rangle $ such that
initialization is not needed as in \cite{DiFranco}.  
However, for simplicity and without loss of
generality, we will assume the system is in the state
$|\mathbf{0}\rangle$.

The Hamiltonian ~(\ref{BHH}) does not include the on-site repulsion $%
H_{U}=\sum n_{j,k}(n_{j,k}-1).$ When the interaction is relatively strong
and/or the total boson umber $\sum n_{j,k}$ is much smaller than the number of
lattice sites, the system tends to have at most one atom at each site due to
the energy gap from $H_{U}$ and when there is only one
particle, a function can be transferred perfectly. In this case, an
arbitrary state $|\phi \>$ of the whole system can be annihilated by the
on-site repulsion Hamiltonian $H_{U}|\phi \>=0.$ Therefore, this case
is equivalent to the one in Eq.~(\ref{N5}) and the transfer will be perfect if we
are able to prepare an initial state $|\phi \>_{1,1}=\alpha |\mathbf{0}\rangle
+\beta b_{1,1}^{\dagger }|\mathbf{0}\rangle$, where only the first site is
occupied. Thus, the state can be perfectly transferred to $\left\vert \phi
\right\>_{M,N} =\alpha |\mathbf{0}\rangle +\beta b_{M,N}^{\dagger }\left\vert
\mathbf{0}\right\rangle $ as is done in the linear case.  Note 
that the result is also applicable to the case where $f(x_{j,k})$ is a
multivariable function, where $j$ and $k$ each range over some number of
sites.

\emph{Perfect state transfer of spin qubits between sites.---} While the above
results are directly applicable to so-called spinless fermions $%
c_{j,k}^{\dagger }$, by replacement $b_{j,k}^{\dagger }\Leftrightarrow
c_{j,k}^{\dagger }$, they can be generalized to realistic cold
fermionic atoms $c_{j,k,\sigma }^{\dagger }$ (or a spherical tensor $%
A_{l_{1}m_{1},l_{2}m_{2},s\sigma }^{\dagger }$) at site $(j,k)$ with a total
spin $s$ and components $\sigma$. Here we consider the example of
$s=1/2$.  One can engineer a spin-independent Hamiltonian similar to 
Eq.~(\ref{BHH}) with quasi-angular momenta in two directions,%
\begin{eqnarray*}
L_{x}^{(1)} &=&\sum_{jk\sigma }C_{j}^{(1)}(c_{j,k,\sigma }^{\dagger
}c_{j+1,k,\sigma }+c_{j+1,k,\sigma }^{\dagger }c_{j,k,\sigma }) \\
L_{x}^{(2)} &=&\sum_{ik\sigma }C_{k}^{(2)}(c_{j,k,\sigma }^{\dagger
}c_{j,k+1,\sigma }+c_{j,k+1,\sigma }^{\dagger }c_{j,k,\sigma })
\end{eqnarray*}%
where the sum is over up and down spins, $\sigma
=\downarrow,\uparrow$. The same construction applies 
to other components of the angular momenta. The most general state 
that can be prepared at site $(1,1)$ is $\left\vert \phi \right\> 
_{1,1}=(\alpha +\beta c_{1,1,\uparrow }^{\dagger }+\gamma c_{1,1,\downarrow
}^{\dagger }+\delta c_{1,1,\uparrow }^{\dagger }c_{1,1,\downarrow }^{\dagger
})\left\vert\mathbf{0}\right\rangle.$ This state can be transferred
perfectly using 
$U(t_{0})\left\vert \phi \right\>_{1,1}=\left\vert\phi\right\>_{M,N}$. 
We can also define the spin operators at a given site in terms
of the Pauli matrices $\vec{\sigma}=(X,Y,Z),$ $\vec{S}%
_{jk}=\sum_{\sigma \sigma ^{\prime }}\left\langle \sigma ^{\prime
}\right\vert \frac{\vec{\sigma }}{2}\left\vert \sigma \right\rangle
c_{j,k,\sigma ^{\prime }}^{\dagger }c_{j,k,\sigma }$. The total spin 
operator is then 
\begin{equation}
\vec{S}=\sum_{jk}\vec{S}_{jk}  \label{spin}
\end{equation}%
which commutes with the quasi-angular momenta $\vec{L}^{(1)}$ and 
$\vec{L}^{(2)}$ and can generate an arbitrary gate on $\left\vert \phi
\right\>_{1,1}$.  

The two possible directions of the spin, $\pm$1/2 
can also represent a spin qubit at site $(j,k)$ when the 
spin is an electron spin in two-dimensional quantum dot 
\cite{Loss:98} or when it is associated with the two states of 
fermionic atoms in an optical lattice \cite{Brennen:99,Sorensen}. 
The states of the qubit are defined by 
$\left\vert 0\right\rangle _{j,k}=c_{j,k,\uparrow }^{\dagger }|\mathbf{0}%
\rangle $ and $\left\vert 1\right\rangle _{j,k}=c_{j,k,\downarrow }^{\dagger
}|\mathbf{0}\rangle$.  An arbitrary state $\left\vert \psi \right\rangle
_{j,k}=\beta \left\vert 0\right\rangle_{j,k}+\gamma \left\vert
1\right\rangle _{j,k}$ of this qubit can be perfectly transferred to $%
\left\vert \psi \right\rangle _{M-j+1,N-k+1}$ by $U(t_{0})$. Any 
generator $\vec{S}$ can be obtained this way so that any rotation to
the state can be constructed. (See the analogue in Eq.~(\ref{Ut}).)  
Also, with the same method an entangled 
state in several sites can be transferred perfectly. For example, an
entangled state $\beta \left\vert 0\right\rangle _{1,1}\left\vert
0\right\rangle _{1,2}+\gamma \left\vert 1\right\rangle _{1,1}\left\vert
1\right\rangle _{1,2}$ is transferred to $\beta \left\vert 0\right\rangle
_{M,N}\left\vert 0\right\rangle _{M,N-1}+\gamma \left\vert 1\right\rangle
_{M,N}\left\vert 1\right\rangle _{M,N-1}$ after time $t_{0}$. 
It should be emphasized that since the possibility of perfect state transfer 
in a spin network of higher dimension remains unclear, the perfect transfer 
via fermionic network (or media) is a promising solution when transferring spin qubits 
in higher dimensions.


\emph{Generalization.---} An arbitrary time-independent unitary
transformation $W$ does not change the commutation relations among angular
momentum components if it corresponds to a similarity transformation
in the Heisenberg picture. Indeed, the whole family of Hamiltonians
generated by an arbitrary $W$ can transfer functions
perfectly. However, this map can introduce new effects. For a spin
system, this transformation $W$ corresponds to the so-called
\emph{dressed qubit}~\cite{Wu/Lidar:03b}. The bosonic Hilbert space is
infinite-dimensional and thus allows more flexibility for 
transformations, including continuous-variable
transformations (which is not possible in spin chains). 

As a simple example, let us first consider transformations $W=\exp (-i\theta
(L_{z}^{(1)}+L_{z}^{(2)}))$. Under the dressing transformation, the
Hamiltonian $H_{l}=J(L_{x}^{(1)}+L_{x}^{(2)})$ becomes 
\[
H^{\prime }=WHW^{\dagger }=\cos \theta H+\sin \theta
J(L_{y}^{(1)}+L_{y}^{(2)})
\]%
The \emph{dressed state} $|\theta \rangle =W|\mathbf{0}\rangle $
allows perfect transfer of any function $f$ using $U(t_{0})$.  The
function is transferred via the coherent state $|\theta \rangle$ as: 
\begin{eqnarray}
&&\hspace{-1.5cm}f(e^{-i(M+N+1)\theta }b_{1,1}^{\dagger })|\theta \rangle  
\nonumber \\
&\rightarrow &f(e^{-i(M+N+1)\theta }b_{M,N}^{\dagger })|\theta \rangle .
\label{coherent}
\end{eqnarray}%

Another example is a dressing transformation which entangles an
individual site $(j,k)$ and a collective bath, 
$W_{jk}=\exp [\lambda(b_{jk}-b_{jk}^{\dagger})B]$ and $W=\prod
W_{jk}$, where $B$ is an operator of the collective bath. Here
$\lambda$ is a small constant. Consider the effect of this
transformation to first order in $\lambda$. The dressed Hamiltonian is 
\[
H^{\prime }=W(H+H_{B})W^{\dagger }=H+H_{B}+\lambda V
\]%
where $V=\sum_{jk}(b_{jk}^{\dagger }B_{jk}+b_{jk}B_{jk}^{\dagger })$, $H_{B}$
is the bath Hamiltonian, and the operators 
$B_{jk}=(J_{j}+J_{j-1}+J_{k}+J_{k-1})B+[H_{B},B]$.  This 
introduces a weak system-bath interaction with a specific $V$ via the 
dressing transformation. Clearly it is possible to introduce different types
of interactions using different types of dressing transformations.
The perfect state transfer from site $(1,1)$ to $(N,M)$ may be roughly
described by $W_{11}|\phi \rangle _{1,1}\left\vert B\right\rangle
\longrightarrow $ $W_{NM}|\phi \rangle _{N,M}\left\vert
 B\right\rangle$, 
where $\left\vert B\right\rangle $ is an eigenstate of the bath
Hamiltonian $H_{B}$.  Even when we assume the bath is in a thermal
equilibrium state, as is usual, the final result is not affected.  

\emph{Perfect state transfer in three dimensional cube.---} In this case, a
particle $b_{j,k,n}^{\dagger }$ (or fermion $c_{j,k,n,\sigma }^{\dagger }$)
at site $(j,k,n)$ is described by a spherical tensor $%
A_{l_{1}m_{1},l_{2}m_{2},l_{3}m_{3}}^{\dagger }.$ While $l_{1}$ and $l_{2}$
are the same as in the two-dimensional case, the third direction is
characterized by $l_{3}=\frac{K-1}{2}$ and $m_{3}=-\frac{K+1}{2}+n.$
The quasi angular momentum $\vec{L}^{(3)}$ in the third direction is
defined analogous to those in (\ref{N11}), for instance, $L_{x}^{(3)}=%
\sum_{jk}C_{j}^{(1)}(b_{j,k,n}^{\dagger }b_{j,k,n+1}+b_{j,k,n+1}^{\dagger
}b_{j,k,})$. The other components are written by replacing $%
b_{j,k,n}^{\dagger }\rightarrow b_{j,k}^{\dagger }$. The evolution operator
is 
\[
U(t)=\exp [-i(J^{(1)}L_{x}^{(1)}+J^{(2)}L_{x}^{(2)}+J^{(3)}L_{x}^{(3)})t]. 
\]%
When $J^{(1)}=J^{(2)}=J^{(3)}=J$ and $t_{0}=\pi /J,$ this expression reduces
to the simple form: 
\begin{equation}
U^{\dagger }(t_{0})\;b_{j,k,n}^{\dagger
}\;U(t_{0})=r_{1}r_{2}r_{3}b_{M-j+1,N-k+1}^{\dagger }\,,  \label{3d}
\end{equation}%
where $r_{3}=\exp \left( -i\pi \frac{K-1}{2}\right).$  All results in two
dimension are directly applicable to the three-dimensional, or even
higher dimensional problems.


\emph{Conclusion.---} We have shown that arbitrary functions can be sent
perfectly (without state initialization or remote collaboration)
through engineered interacting bosonic and fermionic square or 
cube. As an example, we have analyzed the transfer using ultra-cold
bosonic atoms in optical lattices, described by the Bose-Hubbard model
with properly designed site-dependent tunneling amplitudes. In a more
general case, we have studied a family of linear and nonlinear
Hamiltonians that enable perfect state transfers according to dressing
transformations, where a certain noisy factors are considered via the
type of transformations.  As an important consequence, we have shown
that information in spin qubits can be perfectly transferred through
fermionic lattices.  

The ability to send an encoded state perfectly between points, even in
principle, is an important discovery.  For example, one could encode
in a decoherence-free subspace or subsystem to avoid errors
\cite{Zanardi:97c,Duan:98,Lidar:PRL98,Knill:99a,Kempe:00,Lidar:00a}.
(For reviews see \cite{Lidar/Whaley:03,Byrd/etal:pqe04}.)   If there
were imperfections, it has been shown that there exist states which
can transfer quite well even if it is not a PST as discussed in the
introduction.  The ability cancel errors arising from a bath using the
dressing transformations provides yet another useful tool.  This
implies, for example, that physically reasonable global transformations
could be used to cancel some errors.  Taken together, this work
provides methods for robust transfer of an arbitrary function through
two and three dimensional chains which was, until now, not known to
exist.  Furthermore, we have given a Hamiltonian
applicable to a variety of physical systems, leading immediately to
experimental proposals for testing perfect state transfer.

\noindent \emph{Acknowledgments.} This work was supported by the
Ikerbasque Foundation Start-up, the Spanish MEC No.
FIS2009-12773-C02-02, NSF-Grant No. 0545798 to MSB and RGC grant of Hong
Kong No. 7044/08P.


\end{document}